\documentclass{aa}

\usepackage{graphicx}
\usepackage{txfonts}
\usepackage{natbib}

\bibpunct{(}{)}{;}{a}{}{,} % to follow the A&A style

\newcommand{\matrixtt}[4]{\left( \begin{array}{cc}#1&#2\\#3&#4\\\end{array} \right)}

\newcommand{\herm}{H}
\newcommand{\jones}[2]{\vec {#1}_{#2}}
\newcommand{\jonesinv}[2]{\vec {#1}^{-1}_{#2}}
\newcommand{\jonesT}[2]{\vec {#1}^{\herm}_{#2}}
\newcommand{\coh}[2]{\mathsf{{#1}}_{{#2}}}

\begin{document}

\title{Revisiting the radio interferometer measurement equation.\\I. A full-sky Jones formalism}

\author{O.M.\ Smirnov}

\institute{Netherlands Institute for Radio Astronomy (ASTRON)\\
  P.O. Box 2, 7990AA Dwingeloo, The Netherlands \\
  \email{smirnov@astron.nl}}

\date{Received 5 Nov 2010 / Accepted 5 Jan 2011}

\titlerunning{Revisiting the RIME. I. A full-sky Jones formalism}
\authorrunning{O.M.\ Smirnov}

\abstract%
%optional context
{Since its formulation by Hamaker et al., the radio interferometer measurement equation (RIME) 
has provided a rigorous mathematical basis for the development of novel calibration methods and 
techniques, including various approaches to the problem of direction-dependent effects (DDEs). 
However, acceptance of the RIME in the radio astronomical community at large has been slow, which is
partially due to the limited availability of software to exploit its power, and the sparsity of 
practical results. This needs to change urgently.}
%aims
{This series of papers aims to place recent developments in the treatment of DDEs into one 
RIME-based mathematical  framework, and to demonstrate the ease with which the various effects 
can be described and understood. It also aims to show the benefits of a RIME-based approach to calibration.
}
%methods
{Paper I re-derives the RIME from first principles, extends the formalism to
the full-sky case, and incorporates DDEs. Paper II then uses the formalism to describe self-calibration, both with 
a full RIME, and with the approximate equations of older software packages, and shows how this is affected 
by DDEs. It also gives an overview of real-life DDEs and proposed methods of dealing with them. Finally, in Paper III 
some of these methods are exercised to achieve an extremely high-dynamic range calibration of WSRT observations of 3C 147 at 21 cm, 
with full treatment of DDEs.
}%
%results
{The RIME formalism is extended to the full-sky case (Paper I), and is shown to be an elegant way of describing calibration and DDEs (Paper II).
Applying this to WSRT data (Paper III) results in a noise-limited image of the field around 3C 147 with a very high dynamic range (1.6 million), and 
none of the off-axis artifacts that plague regular selfcal. The resulting differential gain solutions contain significant 
information on DDEs and errors in the sky model. 
}%
%optional conclusions
{The RIME is a powerful formalism for describing radio interferometry, and underpins the development of novel calibration methods,
in particular those dealing with DDEs. One of these is the differential gains approach used for the 3C 147 reduction. Differential 
gains can eliminate DDE-related artifacts, and provide information for iterative improvements of sky models. 
Perhaps most importantly, sources as faint as 2 mJy have been shown to yield meaningful differential gain solutions, 
and thus can be used as potential calibration beacons in other DDE-related schemes.}


\maketitle

\section*{Introduction to the series}

The Measurement Equation of a generic radio interferometer (henceforth referred to as the RIME) was formulated by \citet{ME1} after almost 50 years of radio astronomy. Prior to the RIME, mathematical models of radio interferometers (as implemented by a number of software packages such as AIPS, Miriad, NEWSTAR, DIFMAP) were somewhat \emph{ad hoc} and approximate. Despite this (and in part thanks to the careful design of existing instruments), the technique of {\em self-calibration} \citep{Cornwell:selfcal} has allowed radio astronomers to achieve spectacular results. However, by the time the RIME was formulated, even older and well-understood instruments such as the Westerbork Synthesis Radio Telescope (WSRT) and the Very Large Array (VLA) were beginning to expose the limitations of these approximate models. New instruments (and upgrades of older observatories), such as the current crop of Square Kilometer Array \citep{Schilizzi:SKA} ``pathfinders'', and indeed the SKA itself, were already beginning to loom on the horizon. These new instruments exhibit far more subtle and elaborate observational effects, due not only to their greatly increased sensitivity, but also to new features of their design. In particular, while traditional selfcal only deals with direction-independent effects (DIEs), calibration of these new instruments requires us to deal with direction-dependent effects (DDEs), or effects that vary across the field of view (FoV) of the instrument. Following \citet{meqtrees}, I shall refer to \emph{generations} of calibration methods, with first-generation calibration (1GC) predating selfcal, 2GC being traditional selfcal as implemented by the aforementioned packages, and 3GC corresponding to the burgeoning field of DDE-related methods and algorithms. 

It is indeed quite fortunate that the emergence of the RIME formalism has provided us with a complete and elegant mathematical framework for dealing with observational effects, and ultimately DDEs. Oddly enough, outside of a small community of algorithm developers that have enthusiastically accepted the formalism and put it to good use, uptake of RIME by radio astronomers at large has been slow. Even more worryingly, almost 15 years after the first publication, the formalism is hardly ever taught to the new generation of students. This is worrying, because in my estimation, the RIME should be the cornerstone of every entry-level interferometry course! In part, this slow acceptance has been shaped by the availability of software. Today's radio astronomers rely almost exclusively on the 2GC software packages mentioned above, whose internal paradigms are rooted in the selfcal developments of the 1980s and lack an explicit RIME.\footnote{All 2GC packages do use some specific and limited form of the RIME implicitly. This will be discussed further in Paper II \citep{RRIME2}.} On the other hand, relatively few observations were really sensitive enough to push the limits of (or have their science goals compromised by) 2GC. The continued success of legacy packages has meant that the {\em thinking} about interferometry and calibration has still been largely shaped by pre-RIME paradigms. What has not helped this situation is that new software exploiting the power of the RIME has been slow to emerge, and practical results even more so -- but see Paper III \citep{RRIME3} of this series.  

On the other hand, from my personal experience of teaching the RIME at several workshops, once the penny drops, people tend to describe it in terms such as ``obvious'', ``simple'', ``intuitive'', ``elegant'' and ``powerful''. This points at an explanatory gap in the literature. Paper~I of this series therefore tries to address this gap, recasting existing ideas into one consistent mathematical framework, and showing where other approaches to the RIME fit in. It first revisits the ideas of the original RIME papers \citep{ME1,ME4}, deriving the RIME from first principles. It then demonstrates how the fundamentals of interferometry itself (and the van Cittert-Zernike theorem in particular) follow from the RIME (rather than the other way around!), in the process showing how the formalism can incorporate DDEs. This section also looks at alternative formulations of the RIME and their practical implications, and shows where they fit into the formalism. It also tries to clear up some controversies and misunderstandings that have accumulated over the years. Paper~II \citep{RRIME2} then discusses calibration in RIME terms, and explicates the links between the RIME and 2GC implementations of selfcal. 

Paper~II also discusses the subject of DDEs, and places existing approaches into the mathematical framework developed in the preceding sections. DDEs were outside the scope of the original RIME publications, but various authors have been incorporating them into the RIME since. \citet{Rau:DDEs} and \citet{SB:calibration-low-freq} provide an in-depth review of these developments, especially as pertaining to imaging and deconvolution. The above authors have developed a description of DDEs using the $4\times4$ Mueller matrix and coherency vector formalism of the first RIME paper by \citet{ME1}. The $4\times4$ formalism has also been included in the 2nd edition of \citet*[Sect.~4.8]{tms}. In the meantime, \citet{ME4} has recast the RIME using only $2\times2$ matrices. The $2\times2$ form of the RIME has far more intuitive appeal,\footnote{This (admittedly subjective) judgment is firmly based on personal experience of teaching the RIME.} and is far better suited for describing calibration problems, yet has been somewhat unjustly ignored in the literature. Addressing this perceived injustice is yet another aim of these papers. (Section~\ref{sec:formulations} describes the $4\times4$ vs. $2\times2$ formalisms in more detail.)

Last but certainly not least, Paper III \citep{RRIME3} shows an application of these concepts to real data. It presents a record dynamic range (over 1.6 million) calibration of a WSRT observation, including calibration of DDEs. It then analyzes the results of this calibration, shows how the calibration solutions can be used to improve sky models, and demonstrates a rather important implication for the calibratability of future telescopes.

\section{The RIME of a single source\label{sec:me-single-source}}
\label{sec:derivation}

Like many crucial insights, the RIME seems perfectly obvious and simple in hindsight. In fact, it can be almost trivially derived from basic considerations of signal propagation, as shown by \citet{ME1}. In this paper, I will essentially repeat and elaborate on this derivation. This is not original work, but there are several good reasons for reiterating the full argument, as opposed to simply referring back to the original RIME papers. Firstly, some aspects of the basic RIME noted here are not covered by the original papers at all. These are the commutation considerations of Sect.~\ref{sec:taxonomy}, the fact that Jones matrices and coherency matrices behave differently under coordinate transforms (for which reason I even propose a different typographical convention for them), as discussed in Sect.~\ref{sec:circular}, and the 1/2-vs.-1 controversy of Sect.~\ref{sec:factor2}. Then there's the fact that the $2\times2$ version of the formalism proposed by \citet{ME4} and and employed here provides for a much clearer and more intuitive picture that the original $4\times4$ derivation (see Sect.~\ref{sec:mueller} for a discussion), and so deserves far more exposure in the literature than the sole Hamaker paper to date. Finally, I want to establish some typographical conventions and mathematical nomenclature, and lay the groundwork for my own extensions of the formalism, which start at Sect.~\ref{sec:full-sky-rime}. This seemed sufficient reason to give a complete derivation of the RIME from scratch.

In Sects.~\ref{sec:me-multiple-sources} and \ref{sec:full-sky-rime}, I extend the $2\times2$ formalism into the image-plane domain, show how the van Cittert-Zernike (VCZ) theorem naturally follows from the RIME, and sketch the problem of DDEs. Section \ref{sec:closures} elaborates some RIME-based closure relationships, Sect.~\ref{sec:rime-limitations} then examines some important limitations and boundaries of the RIME formalism, and Sect.~\ref{sec:formulations} looks at alternative formulations of the RIME. Finally, Sect.~\ref{sec:controversies} attempts to clear up some errors and controversies surrounding the formalism.

\subsection{Signal propagation}

Consider a single source of quasi-monochromatic signal (i.e. a sky consisting of a single point source). The signal at a fixed point in space and time can be then be described by the complex vector $\vec e$. Let us pick an orthonormal $xyz$ coordinate system, with $z$ along the direction of propagation (i.e. from antenna to source). In such a system, $\vec e$ can be represented by a column vector of 2 complex numbers:

\[
\vec e = \left( \begin{array}{c}e_x\\e_y\end{array} \right) 
\]

Our fundamental assumption is {\em linearity}: all transformations along the signal path are linear w.r.t. $\vec e$. Basic linear algebra tells us that all linear transformations of a 2-vector can be represented (in any given coordinate system) by a matrix multiplication:

\[
\vec e' = \jones{J}{} \vec e,
\]

where $\jones{J}{}$ is a $2\times2$ complex matrix known as the {\em Jones} matrix \citep{jones}. Obviously, multiple effects along the signal propagation path correspond to repeated matrix multiplications, forming what I call a {\em Jones chain}. We can regard multiple effects separately and write out Jones chains, or we can collapse them all into a single cumulative Jones matrix as convenient:

\begin{equation}\label{eq:jones-chain}
\vec e' = \jones{J}{n} \jones{J}{n-1} ... \jones{J}{1} \vec e = \jones{J}{} \vec e
\end{equation}

The order of terms in a Jones chain corresponds to the physical order in which the effects occur along the signal path. Since matrix multiplication does not (in general) commute, we must be careful to preserve this order in our equations.

Now, the signal hits our antenna and is ultimately converted into complex voltages by the antenna feeds. Let us further assume that we have two feeds $a$ and $b$ (for example, two linear dipoles, or left/right circular feeds), and that the voltages $v_a$ and $v_b$ are linear w.r.t. $\vec e$. We can formally treat the two voltages as a voltage vector $\vec v$, analogous to $\vec e$. Their linear relationship is yet another matrix multiplication:

\begin{equation}\label{eq:e-voltage}
\vec v = \left( \begin{array}{c}v_a\\v_b\end{array} \right) = \jones{J}{} \vec e
\end{equation}
 
Equation~(\ref{eq:e-voltage}) can be thought of as representing the fundamental linear relationship between the voltage vector $\vec v$ as measured by the antenna feeds, and the ``original'' signal vector $\vec e$ at some arbitrarily distant point, with $\jones{J}{}$ being the cumulative product of all propagation effects along the signal path (including electronic effects in the antenna/feed itself). I shall call refer to this $\jones{J}{}$ as the {\em total Jones} matrix, as distinct from the individual Jones terms in a Jones chain.

\subsection{The visibility matrix}

Two spatially separated antennas $p$ and $q$ measure two independent voltage vectors $\vec v_p,\vec v_q$. In an {\em interferometer}, these are fed into a correlator, which produces 4 pairwise correlations between the components of $\vec v_p$ and $\vec v_q$:

    \begin{equation}\label{eq:correlation}
    \langle v_{pa}v^*_{qa}\rangle, \langle v_{pa}v^*_{qb}\rangle, 
    \langle v_{pb}v^*_{qa}\rangle, \langle v_{pb}v^*_{qb}\rangle
    \end{equation}

Here, angle brackets denote averaging over some (small) time and frequency bin, and $x^*$ is the complex conjugate of $x$.  It is convenient for our purposes to arrange these four correlations into the {\em visibility matrix\/}\footnote{\citet{ME4} calls $\coh{V}{pq}$ the {\em coherency matrix}, in order to distinguish it from traditional scalar visibilities. Since the elements of the matrix are precisely the complex visibilities, I submit {\em visibility} matrix as a more logical term.} $\coh{V}{pq}$:

    \[
    \coh{V}{pq} = 2 \matrixtt{\langle v_{pa}v^*_{qa}\rangle}{\langle v_{pa}v^*_{qb}\rangle}{\langle v_{pb}v^*_{qa}\rangle}{\langle v_{pb}v^*_{qb}\rangle}
    \]

I introduce a factor of 2 here, for reasons explained in Sect.~\ref{sec:factor2}. It is easily seen that $\coh{V}{pq}$ can be written as a matrix product of $\vec v_p$ (as a column vector), and the conjugate of $\vec v_q$ (as a row vector):

\begin{equation}\label{eq:coherency}
\coh{V}{pq} = 2 \left<\left( \begin{array}{c}v_{pa}\\v_{pb}\end{array} \right) (v^*_{qa},v^*_{qb}) \right > = 2 \langle \vec v_p \vec v^\herm_q \rangle
\end{equation}

Here, $\herm$ represents the conjugate transpose operation (also called a Hermitian transpose).

\subsection{\label{sec:RIME-emerges}The RIME emerges}

Starting with some arbitrarily distant vector $\vec e$, our signal travels along two different paths to antennas $p$ and $q$. Following Eq.~(\ref{eq:e-voltage}), each propagation path has its own total Jones matrix, $\jones{J}{p}$ and $\jones{J}{q}$. Combining Eqs.~(\ref{eq:e-voltage}) and (\ref{eq:coherency}), we get:

    \begin{equation}\label{eq:corr1}
    \coh{V}{pq} = 2 \langle  \jones{J}{p} \vec e ( \jones{J}{q} \vec e )^\herm \rangle  = 2 \langle  \jones{J}{p} (\vec e \vec e^\herm) \jonesT{J}{q} \rangle 
    \end{equation}

Assuming that $\jones{J}{p}$ and $\jones{J}{q}$ are constant over the averaging interval,\footnote{This is a crucial assumption, which I will revisit in Sect.~\ref{sec:smearing}.} we can move them outside the averaging operator:

    \begin{equation}\label{eq:corr2}
    \coh{V}{pq} = 2 \jones{J}{p} \langle  \vec e \vec e^\herm \rangle  \jonesT{J}{q} = 
    2 \jones{J}{p} 
    \matrixtt{\langle e_x e^*_x\rangle }{\langle e_x e^*_y\rangle }{\langle e_y e^*_x\rangle }{\langle e_y e^*_y\rangle }
    \jonesT{J}{q}
    \end{equation}

The bracketed quantities here are intimately related to the definition of the Stokes parameters \citep{born-wolf,tms}. \citet{ME3} explicitly show that

    \begin{equation}\label{eq:IQUV}
    2 
    \matrixtt{\langle e_x e^*_x\rangle }{\langle e_x e^*_y\rangle }{\langle e_y e^*_x\rangle }{\langle e_y e^*_y\rangle }
    = 
    \matrixtt{I+Q}{U+iV}{U-iV}{I-Q} = \coh{B}{}
    \end{equation}

I now define the {\em brightness matrix} $\coh{B}{}$ as the right-hand side\footnote{Following a long-standing controversy, I have decided to break with \citet{ME4} by omitting $\frac{1}{2}$ from the definition of $\coh{B}{}$, and adding a factor 2 to the definition of $\coh{V}{pq}$ in Eq.~(\ref{eq:coherency}). The reasons for this will be spelled out in Sect.~\ref{sec:factor2}.} of Eq.~(\ref{eq:IQUV}). This gives us the first form of the RIME, that of a single point source:

    \begin{equation}\label{eq:me0}
    \coh{V}{pq} = \jones{J}{p} \coh{B}{}  \jonesT{J}{q}
    \end{equation}

Or in expanded form:

\[
    \left( 
    \begin{array}{cc}
    v_{aa} & v_{ab} \\
    v_{ba} & v_{bb} \\
    \end{array}
    \right) = 
    \left( 
    \begin{array}{cc}
    j_{11p} & j_{12p} \\
    j_{21p} & j_{22p} \\
    \end{array}
    \right) 
    \left( 
    \begin{array}{cc}
    I+Q & U+iV \\
    U-iV & I-Q \\
    \end{array}
    \right) 
    \left( 
    \begin{array}{cc}
    j_{11q} & j_{12q} \\
    j_{21q} & j_{22q} \\
    \end{array}
    \right)^\herm
\]

which quite elegantly ties together the observed visibilities $\coh{V}{pq}$ with the intrinsic source brightness $\coh{B}{}$, and the per-antenna terms $\jones{J}{p}$ and $\jones{J}{q}$.

Note that Eq.~(\ref{eq:me0}) holds in any coordinate system. The vector $\vec e$, the brightness matrix $\coh{B}{}$ that is derived from it, and the linear transformations $\jones{J}{p}$ and $\jones{J}{q}$ are distinct mathematical entities that are independent of coordinate systems; choosing a coordinate basis associates a specific {\em representation} with $\vec e$,  $\coh{B}{}$ and $\jones{J}{}$, manifesting itself in a 2-vector or a $2\times2$ matrix populated with specific complex numbers. For example, it is quite possible (and sometimes desirable) to rewrite the RIME in a circular polarization basis. This is discussed further in Sect.~\ref{sec:circular}. In this paper, I shall use an orthonormal $xyz$ basis unless otherwise stated.

\subsection{Some typographical conventions}

Throughout this series of papers, I shall adopt the following typographical conventions for formulas:

\begin{description}
\item[Scalar quantities] will be indicated by lower- and uppercase italics: $e_x,I,K_p$.
\item[Vectors] will be indicated by lowercase bold italics: $\vec e$.
\item[Jones matrices] will be indicated by uppercase bold italics: $\jones{J}{}$. As a special case, scalar matrices
(Sect.~\ref{sec:taxonomy}) will be indicated by normal-weight italics: $K_p$.
\item[Visibility, coherency and brightness matrices] will be indicated by sans-serif font: 
$\coh{B}{}, \coh{V}{pq}, \coh{X}{pq}$. This emphasizes their different mathematical nature (and in particular, that they transform differently under change of coordinate frame, Sect.~\ref{sec:circular}).
\end{description}

\subsection{The ``onion'' form}

We can also choose to expand $\jones{J}{p}$ and $\jones{J}{q}$ into their associated Jones chains, as per 
Eq.~(\ref{eq:jones-chain}). This results in the rather pleasing ``onion'' form of the RIME:

    \begin{equation}\label{eq:me0-onion}
    \coh{V}{pq} = \jones{J}{pn}(...(\jones{J}{p2} (\jones{J}{p1} \coh{B}{}  \jones{J}{q1}^\herm)\jonesT{J}{q2}) ... )\jonesT{J}{qm}
    \end{equation}

Intuitively, this corresponds to various effects in the signal path applying sequential layers of ``corruptions'' to the original source brightness $\coh{B}{}$. Note that the two signal paths can in principle be entirely dissimilar, making the ``onion'' asymmetric (hence the use of $n\ne m$ for the outer indices). An example of this is VLBI with \emph{ad hoc} arrays composed of different types of telescopes. One of the strengths of the RIME is its ability to describe heterogeneous interferometer arrays with dissimilar signal propagation paths.

\subsection{An elementary Jones taxonomy\label{sec:taxonomy}}

Different propagation effects are described by different kinds of Jones matrices. The simplest kind of matrix is a {\em scalar} matrix, corresponding to a transformation that affects both components of the $\vec e$ vector equally. I shall use normal-weight italics $(K)$ to emphasize scalar matrices. An example is the phase delay matrix below:

    \[
    K = {\rm e}^{i\phi} \equiv 
    \left( 
    \begin{array}{cc}
    {\rm e}^{i\phi} & 0 \\
    0 & {\rm e}^{i\phi} \\
    \end{array}
    \right) =   
    {\rm e}^{i\phi} \left( 
    \begin{array}{cc}
    1 & 0 \\
    0 & 1 \\
    \end{array}
    \right)    
    \]

An important property of scalar matrices is that they have the same representation in all coordinate systems, so {\em scalarity} is defined independently of coordinate frame.

Diagonal matrices correspond to effects that affect the two $\vec e$ components independently, without intermixing. Note that unlike scalarness, diagonality {\em does} depend on choice of coordinate systems. For example, if we consider linear dipoles, their electronic gains are (nominally) independent, and the corresponding Jones matrix is diagonal in an $xy$ coordinate basis:

    \[
    \jones{G}{} = 
    \left( 
    \begin{array}{cc}
    g_x & 0 \\
    0 & g_y \\
    \end{array}
    \right) 
    \]

The gains of a pair of circular receptors, on the other hand, are not diagonal in an $xy$ frame (but are diagonal in a circular polarization frame -- see Sect.~\ref{sec:circular}).

Matrices with non-zero off-diagonal terms intermix the two components of $\vec e$. A special case of this is the {\em rotation} matrix:

    \[
    \mbox{Rot~}\phi = 
    \left( 
    \begin{array}{cc}
    \cos\phi & -\sin\phi \\
    \sin\phi & \cos\phi \\
    \end{array}
    \right) 
    \]

Like diagonality, the property of being a rotation matrix also depends on choice of coordinate frame. Examples of rotation matrices (in an $xy$ frame) are rotation through parallactic angle $\jones{P}{}$, and Faraday rotation in the ionosphere $\jones{F}{}$. Note also that rotation in an $xy$ frame becomes a special kind of diagonal matrix in the circular frame (see Sect.~\ref{sec:circular}).

It is important for our purposes that, while in general matrix multiplication is non-commutative, specific kinds of matrices do commute:

\begin{enumerate}
\item Scalar matrices commute with everything.
\item Diagonal matrices commute among themselves.
\item Rotation matrices commute among themselves\footnote{Note that this is only true for $2\times2$ matrices. Higher-order rotations do not commute.}.
\end{enumerate}

Rules 2 and 3 are not very satisfactory as stated, because ``diagonal'' and ``rotation'' are properties defined in a specific coordinate frame, while (non-)commutation is defined independently of coordinates: two linear operators $\jones{A}{}$ and $\jones{B}{}$ either commute or they don't, so their matrix representations must necessarily commute (or not) irrespective of what they look like for a particular basis. Let us adopt a practical generalization: 

\paragraph{The Commutation Rule:} if there exists a coordinate basis in which $\jones{A}{}$ and $\jones{B}{}$ are both diagonal (or both a rotation\footnote{As noted above, rotation can become diagonality through change of coordinate basis, so this doesn't actually add anything to our general rule.}), then $\jones{A}{} \jones{B}{}=\jones{B}{}\jones{A}{}$ in all coordinate frames. 

We shall be making use of commutation properties later on.

\subsection{\label{sec:coherency}Phase and coherency}

Equation~(\ref{eq:me0}) is universal in the sense that the $\jones{J}{p}$ and $\jones{J}{q}$ terms represent all effects along the signal path rolled up into one $2\times2$ matrix. It is time to examine these in more detail. In the ideal case of a completely uncorrupted observation, there is one fundamental effect remaining -- that of phase delay associated with signal propagation. We are not interested in absolute phase, since the averaging operator implicit in a correlation measurement such as Eq.~(\ref{eq:correlation}) is only sensitive to phase {\em difference} between voltages $\vec v_p$ and $\vec v_q$. 

Phase difference is due to the geometric pathlength difference from source to antennas $p$ and $q$. For reasons discussed in Sect.~\ref{sec:smearing}, we want to minimize this difference for a specific direction, so a correlator will usually introduce additional delay terms to compensate for the pathlength difference in the chosen direction, effectively ``steering'' the interferometer. This direction is called the {\em phase centre}. The conventional approach is to consider phase differences on {\em baseline} $pq$, but for our purposes let's pick an arbitrary zero point, and consider the phase difference at each antenna $p$ relative to the zero point.

Let us adopt the conventional coordinate system\footnote{Note that there is some unfortunate confusion in coordinate systems used in radio interferometry. The \citet{IAU74} defines Stokes parameters in a right-handed coordinate system with $x$ and $y$ in the plane of the sky towards North and East, and the $z$ axis pointing towards the observer. The conventional $lm$ frame has $l$ pointing East and $m$ North. In practice, this means that rotation through parallactic angle must be applied in one direction in the $lm$ frame, and in the opposite direction in the polarization frame. The formulations of the present paper are not affected.} and notations \citep[see e.g.][]{tms}, with the $z$ axis pointing towards the phase centre, and consider antenna $p$ located at coordinates $\vec u_p=(u_p,v_p,w_p)$. The phase difference at point $\vec u_p$ relative to $\vec u=0$, for a signal arriving from direction $\vec\sigma$, is given by

  \[
  \kappa_p = 2\pi\lambda^{-1}(u_p l+v_p m+w_p (n-1)),
  \]

where $l,m,n=\sqrt{1-l^2-m^2}$ are the direction cosines of $\vec\sigma$, and $\lambda$ is signal wavelength. It is customary to define $\vec u$ in units of wavelength, which allows us to omit the $\lambda^{-1}$ term.
Following \citet{JEN:note185}, I can now introduce a scalar {\em $K$-Jones} matrix representing the phase delay effect. After all, phase delay is just another linear transformation of the signal, and is perfectly amenable to the Jones formalism:

  \begin{equation}\label{eq:K}
  K_p = {\rm e}^{-i\kappa_p} = {\rm e}^{-2\pi i(u_p l+v_p m+w_p (n-1))}
  \end{equation}

The RIME for a single uncorrupted point source is then simply:

  \begin{equation}\label{eq:me-point-source}
  \coh{V}{pq} = K_p \coh{B}{}  K^\herm_q
  \end{equation}

Substituting the exponents for $K_p$ from Eq.~(\ref{eq:K}), and remembering that scalar matrices commute with everything, we can recast Eq.~(\ref{eq:me-point-source}) in a more traditional form:\footnote{The sign of the exponent in these equations is a matter of convention, and is therefore subject to perennial confusion. WSRT software uses ``$-$'', but has used ``$+$'' in the past. VLA software seems to use ``$+$''. Fortunately, in practice it is usually easy to tell which convention is being used, and conjugate the visibilities if needed.}

  \begin{equation}\label{eq:me-point-source-uvw}
  \coh{V}{pq} = \coh{B}{}  {\rm e}^{-2\pi i(u_{pq} l+v_{pq} m+w_{pq} (n-1))},\;\vec u_{pq} = \vec u_p - \vec u_q,
  \end{equation}
 
which expresses the visibility as a function of {\em baseline $uvw$ coordinates} $\vec u_{pq}$. I shall call the visibility matrix given by Eqs.~(\ref{eq:me-point-source}) or (\ref{eq:me-point-source-uvw}) the {\em source coherency}, and write it as $\coh{X}{pq}$. In the traditional view of radio interferometry, $\coh{X}{pq}$ is a measurement of the coherency function $\coh{X}{}(u,v,w)$ at point $u_{pq},v_{pq},w_{pq}$ (with $\coh{X}{}$ being a $2\times2$ complex matrix rather than the traditional scalar complex function). For the purposes of these papers, let us adopt an operational definition of {\em source coherency} as being the visibility that would be measured by a corruption-free interferometer. For a point source, the coherency is given by Eq.~(\ref{eq:me-point-source}).

\subsection{A single corrupted point source}

A real-world interferometer will have some ``corrupting'' effects in the signal path, in addition to the nominal phase delay $K_p$. Since the latter is scalar and thus commutes with everything, we can move it to the beginning of the Jones chain, and write the total Jones $\jones{J}{p}$ of Eq.~(\ref{eq:me0}) as

\[
\jones{J}{p} = \jones{G}{p} K_p,
\]

where $\jones{G}{p}$ represents all the other (corrupting) effects. We can then formulate the RIME for a single corrupted point source as:

  \begin{equation}\label{eq:me-point-source-corrupted}
  \coh{V}{pq} = \jones{G}{p} \coh{X}{pq} \jonesT{G}{q},
  \end{equation}

where $\coh{X}{pq}$ is the source coherency, as defined above.

\section{Multiple discrete sources\label{sec:me-multiple-sources}}

Let us now consider a sky composed of $N$ point sources. The contributions of each source to the measured visibility matrix $\coh{V}{pq}$ add up linearly. The signal propagation path is different for each source $s$ and antenna $p$, but each path can be described by its own Jones matrix $\jones{J}{sp}$. Equation~(\ref{eq:me0}) then becomes:

  \begin{equation}\label{eq:me-nps-j}
  \coh{V}{pq} = \sum_{s}{\jones{J}{sp} \coh{B}{s} J^\herm_{sq}}
  \end{equation}

Remember that each $\jones{J}{sp}$ is a product of a (generally non-commuting) {\em Jones chain}, corresponding to the physical order of effects along the signal path:

  \[
  \jones{J}{sp} = \jones{J}{spn} ... \jones{J}{sp1},
  \]

where effects represented by the right side of the chain ($...\jones{J}{sp1}$) occur ``at the source'', and effects on the left side of the chain ($\jones{J}{spn}...$) ``at the antenna''. Somewhere along the chain is the phase term $K_{sp}$, but since (being a scalar matrix) it commutes with everything, we are free to move it to any position in the product.

Some elements in the chain may be the same for all sources. This tends to be true for effects at the antenna end of the signal path, such as electronic gain. Let us then collapse the chain into a product of three Jones matrices:

  \[
  \jones{J}{sp} = \jones{G}{p} \jones{E}{sp} K_{sp}
  \]

$\jones{G}{p}$ is the source-independent ``antenna'' (left) side of the Jones chain, i.e. the product of the terms beginning with $\jones{J}{spn}$, up to and not including the leftmost source-dependent term (if the entire chain is source-dependent, $\jones{G}{p}$ is simply unity), $\jones{E}{sp}$ is the source-dependent remainder of the chain, and $K_{sp}$ is the phase term. We can then recast Eq.~(\ref{eq:me-nps-j}) as follows:

  \begin{equation}\label{eq:me-nps-gek}
  \coh{V}{pq} = \jones{G}{p} \left ( \sum_{s}{\jones{E}{sp} K_{sp} \coh{B}{s} K^\herm_{sq} \jonesT{E}{sq}} \right ) \jonesT{G}{q}
  \end{equation}

Or, using the source coherency of Eq.~(\ref{eq:me-point-source}):

  \begin{equation}\label{eq:me-nps-ge}
  \coh{V}{pq} = \jones{G}{p} \left ( \sum_{s}{\jones{E}{sp} \coh{X}{spq} E^\herm_{sq}} \right ) \jonesT{G}{q}
  \end{equation}

$\jones{G}{p}$ describes the {\em direction-independent} effects (DIEs), or the \emph{uv-Jones} terms, and $\jones{E}{sp}$ the {\em  direction-dependent} effects (DDEs), or the \emph{sky-Jones} terms. 

In principle, the sum in Eq.~(\ref{eq:me-nps-ge}) should be taken over all sufficiently bright\footnote{Brighter than the noise, that is -- see Sect.~\ref{sec:noise}.} sources in the sky, but in practice our FoV is limited by the voltage beam pattern of each antenna, or by the horizon, in the case of an all-sky instrument such as the Low Frequency Array (LOFAR). In RIME terms, beam gain is just another Jones term in the chain, ensuring $\jones{E}{sp}\to 0$ for sources outside the beam.

If the observed field has little to none spatially extended emission, this form of the RIME is already powerful enough to allow for calibration of DDEs, as I shall show in Paper III \citep{RRIME3}.

\section{The full-sky RIME\label{sec:full-sky-rime}}

In the more general case, the sky is not a sum of discrete sources, but rather a continuous brightness distribution $\coh{B}{}(\vec\sigma)$, where $\vec\sigma$ is a (unit) direction vector. For each antenna $p$, we then have a Jones term $\jones{J}{p}(\vec\sigma)$, describing the signal path for direction $\vec\sigma$. To get the total visibility as measured by an interferometer, we must integrate Eq.~(\ref{eq:me0}) over all possible directions, i.e. over a unit sphere:

\[
\coh{V}{pq} = \int\limits_{4\pi} \jones{J}{p}(\vec\sigma) \coh{B}{}(\vec\sigma) \jonesT{J}{q}(\vec\sigma) \, d\Omega
\]

This spherical integral is not very tractable, so we perform a sine projection of the sphere onto the plane $(l,m)$ 
tangential at the field centre.\footnote{Or the pole, for East-West arrays, which does not materially change any of the arguments.} Note that this analysis is fully analogous to that of \citet[Sect.~3.1]{tms}, with only the integrand being somewhat different.  The integral then becomes:

\[
\coh{V}{pq} = \iint\limits_{lm} \jones{J}{p}(\vec l) \coh{B}{}(\vec l) \jonesT{J}{q}(\vec l) \frac{dl\,dm}{n},
\;\;\mathrm{where}\; n=\sqrt{1-l^2-m^2}.
\]

I'm going to use $\vec l$ and $(l,m)$ interchangeably from now on. By analogy with Eq.~(\ref{eq:me-nps-gek}), we now decompose $\jones{J}{p}(\vec l)$ into a direction-independent part $\jones{G}{}$, a direction-dependent part $\jones{\bar E}{}$, and the phase term $K$:

\[
\jones{J}{p}(\vec l) = \jones{G}{p}\jones{\bar E}{p}(\vec l) K_p(\vec l) = \jones{G}{p}\jones{\bar E}{p}(\vec l) {\rm e}^{-2\pi i(u_p l+v_p m+w_p (n-1))}
\]

Substituting this into the integral, and commuting the $K$ terms around, we get

\begin{equation}\label{eq:me-allsky0}
\coh{V}{pq} = \jones{G}{p} \left( \iint\limits_{lm} \frac{1}{n} \jones{\bar E}{p} \coh{B}{} \jonesT{\bar E}{q} \mathrm{e} ^{-2\pi i(u_{pq} l+v_{pq} m+w_{pq} (n-1))} \,dl\,dm \right) \jonesT{G}{q}
\end{equation}

This equation is one form of a general full-sky RIME. It is in fact a type of three-dimensional Fourier transform; the \emph{non-coplanarity} term in the exponent, $w_{pq}(n-1)$, is what prevents us from treating it as the much simpler 2D transform. Since $w_{pq}=w_p-w_q$, we can decompose the non-coplanarity term into per-antenna terms $W_p=\frac{1}{\sqrt{n}} \mathrm{e}^{-2\pi i w_p (n-1)}$. These can be thought of direction-dependent Jones matrices in their own right, and subsumed into the overall sky-Jones term by defining $\jones{E}{p} = \jones{\bar E}{p}W_p$. The full-sky RIME (Eq.~\ref{eq:me-allsky0}) can then be rewritten using a 2D Fourier Transform of the \emph{apparent sky as seen by baseline $pq$}, or $\coh{B}{pq}$:

\begin{eqnarray}\label{eq:me-allsky}
\coh{V}{pq} & = & \jones{G}{p} \left( \iint\limits_{lm} \coh{B}{pq} \mathrm{e} ^{-2\pi i(u_{pq} l+v_{pq} m)} \,dl\,dm \right) \jonesT{G}{q}, \\
\nonumber \coh{B}{pq} & \equiv & \jones{E}{p} \coh{B}{} \jones{E}{q}
\end{eqnarray}

I shall return to this general formulation in Paper II \citep{RRIME2}. In the meantime, consider the import of those $pq$ indices in $\coh{B}{pq}$. They are telling us that we're measuring a 2D Fourier Transform of the sky -- but the ``sky'' is different for every baseline! This violates the fundamental premise of traditional selfcal, which assumes that we're measuring the F.T. of one common sky. From the above, it follows that this premise only holds when all DDEs are \emph{identical} across all antennas: $\jones{E}{p}(\vec l) \equiv \jones{E}{}(\vec l)$ (or at least where $\coh{B}{}(\vec l) \ne 0$). Only under this condition does the apparent sky $\coh{B}{pq}$ become the same on all baselines (in the traditional view, this corresponds to the ``true'' sky attenuated by the power beam):

\[
\coh{B}{pq}(\vec l) \equiv \coh{B}{\mathrm{app}}(\vec l) =  \jones{E}{}(\vec l) \coh{B}{}(\vec l) \jonesT{E}{}(\vec l)
\]

If this is met, we can then rewrite the full-sky RIME as:

\begin{equation}\label{eq:me-allsky-simple}
\coh{V}{pq} = \jones{G}{p} \coh{X}{pq} \jonesT{G}{q},
\end{equation}
where $\coh{X}{pq} = \coh{X}{}(u_{pq},v_{pq})$, and the matrix function $\coh{X}{}(\vec u)$ is simply the (element-by-element) two-dimensional Fourier transform\footnote{Note that I'm using $\vec u$ as a shorthand for both $(u,v)$ and $(u,v,w)$, depending on context.}
of the matrix function $\coh{B}{\rm app}(\vec l)$. I shall also write this as $\coh{X}{}={\cal F}\coh{B}{\mathrm{app}}$. The similarity to Eq.~(\ref{eq:me-point-source-corrupted}) of a single point source is readily apparent. For obvious reasons, I shall call $\coh{X}{}(\vec u)$ the {\em sky coherency}. Effectively, we have derived the van Cittert-Zernike theorem (VCZ), the cornerstone of radio interferometry \citep[Sect.~14.1]{tms}, from the basic RIME! 

Such an approach turns the original original coherency matrix formulation of \citet{ME4} on its head. Note that Eq.~(\ref{eq:me-allsky-simple}) here is the same as Eq.~(2) of that work. In the RIME papers, Hamaker et al. defer to VCZ, treating the coherency as a ``given'' (while recasting it to matrix form) to which Jones matrices then apply. Treating phase ($K$) as a Jones matrix in its own right \citep{JEN:note185} allows for a natural extension of the Jones formalism into the $(l,m)$ plane, and shows that VCZ is actually a consequence of the RIME rather than being something extrinsic to it. This also allows DDEs to be incorporated into the same formalism, in a manner similar to that suggested for $w$-projection \citep{Cornwell:wproj}. I shall return to this subject in Paper II \citep{RRIME2}.

\subsection{Time variability and the fundamental assumption of selfcal\label{sec:timevar}}

I have hitherto ignored the time variable. Signal propagation effects, and indeed the sky itself, do vary in time, but the RIME describes an effectively instantaneous measurement (ignoring for the moment the issue of time averaging, which will be considered separately in Sect.~\ref{sec:smearing}). Time begins to play a critical role when we consider DDEs. 

At any point in time, an interferometer given by Eq.~(\ref{eq:me-allsky-simple}) measures the coherency function $\coh{X}{}(\vec u)$ at a number of points $\vec u_{pq}$ (i.e. for all baselines $pq$). This ``snapshot'' measurement gives a limited sampling of the $uv$ plane. To sample the $uv$ plane more fully, we usually rely on the Earth's rotation, which over several hours effectively ``swings'' every baseline vector $\vec u_{pq}$ through an arc in the $uv$ plane. Therefore, for Eq.~(\ref{eq:me-allsky-simple}) to hold throughout an observation, we must additionally assume that the apparent sky $\coh{B}{\rm app}$ remains constant over the observation time! In other words, unless we're dealing with snapshot imaging, the $\jones{E}{p}\equiv\jones{E}{}$ assumption must be further augmented:

\begin{equation}\label{eq:trivial-ddes}
\jones{E}{p}(t,\vec l) \equiv \jones{E}{p}(\vec l) \equiv \jones{E}{}(\vec l)\;\;\mbox{for all~} t,p. 
\end{equation}

This equation captures the fundamental assumption of traditional selfcal. I shall call DDEs that satisfy Eq.~(\ref{eq:trivial-ddes}) \emph{trivial DDEs}. As shown above, trivial DDEs effectively replace the true sky $\coh{B}{}$ by a single apparent sky $\coh{B}{\mathrm{app}}$, and are not usually a problem for calibration, since they can be corrected for entirely in the image plane.\footnote{Even then things are not always easy. Rapid variation in frequency, such as the 17 MHz ``ripple'' of the WSRT primary beam \citep[see Paper II,][Sect.~2.1.1]{RRIME2} can cause considerable difficulty for spectral line calibration, even if the DDE is trivial in the sense of Eq.~(\ref{eq:trivial-ddes}).}  For example, the primary beam gain is usually treated as a trivial DDE in 2GC packages \citep[see Paper II,][Sect.~2.1]{RRIME2}. 

Equation~(\ref{eq:trivial-ddes}) is most readily met with narrow FoVs (i.e. with $\jones{E}{p}$ rapidly going to zero away from the field centre, leaving little scope for other variations), small arrays (small $w_p$, also all stations see through the same atmosphere), higher frequencies (narrow FoV, less ionospheric effects), and also with coplanar arrays such as the WSRT ($w_p\equiv0$, thus $W_p\equiv1$). The new crop of instruments is, of course, trending in the opposite direction on all these points, and is thus subject to far more severe and non-trivial DDEs.

% The traditional way of ensuring that Eq.~(\ref{eq:trivial-ddes}) holds is by observig a narrow FoV
% 
% 
% in some depth. On the one hand, 2GC packages do work rather well, so it must hold most of the time. On the other hand, $\jones{E}{p}$ contains $W_p$, which must break the requirement, being variable in time and per-antenna. There are two common ``easy'' cases:
% 
% {\bf Narrow FoV.} The $\jones{E}{p}$ term includes the antenna beam pattern, which (for traditional dishes, at higher frequencies) ensures that $\jones{E}{p} \to 0$ as we get away from the pointing centre (which is usually the same as phase centre, $l=m=0$). In addition, time averaging (Sect.~\ref{sec:smearing}) also tends to drive the observed $\coh{B}{}$ to zero as we get away from phase centre. Besides, $W_p\to1$ near phase centre, so a narrow FoV usually ensures that the relationship of Eq.~(\ref{eq:trivial-ddes}) is met.
% 
% {\bf Coplanar arrays:} for E-W arrays such as the WSRT, all antennas lie in the same plane, so we can construct a coordinate system with $w_p\equiv0$. 
% 
% $\jones{E}{}(l,m)$ is only non-zero in a small area around $l=m=0$ (so that $n\to 1$ where $E(l,m) \ne 0.$) This is the case in traditional narrow-FoV interferometery. Alternatively, $w\equiv0$ (as is the case in a  coplanar interferometer array, such as an East-West array).

\section{Matrix closures and singularities\label{sec:closures}}

Scalar closure relationships have played an important role in 2GC calibration, both as a diagnostic tool, and as an observable. Traditionally, these are expressed in terms of a three-way phase closure and a four-way amplitude closure \citep[see e.g.][Sect.~10.3]{tms}. Since the underlying premise of a closure relationship is that observed scalar visibilities can be expressed in terms of per-antenna scalar gains, and the RIME is a generalization of the same premise in matrix terms, it seems worthwhile to see if a general matrix (i.e. fully polarimetric) closure relationship can be derived.

Indeed, in the case of a single point source, we can write out a four-way closure for antennas $m,n,p,q$ as follows:

\begin{equation}\label{eq:closure}
\coh{V}{mn}\coh{V}{pn}^{-1}\coh{V}{pq}\coh{V}{mq}^{-1} = 1
\end{equation}

The above equation can be easily verified by substituting in Eq.~(\ref{eq:me0}) for each visibility term, and remembering that $(\jones{A}{}\jones{B}{})^{-1}=\jonesinv{B}{}\jonesinv{A}{}$. 

Since matrix inversion is involved, the essential requirement here is non-singularity of all matrices in Eq.~(\ref{eq:me0}). The brightness matrix $\coh{B}{}$ is non-singular by definition (unless it's trivially zero), but what does it mean for a Jones matrix to be singular? Some examples of singular matrices are:

\[
\matrixtt{a}{0}{0}{0}, \; \matrixtt{a}{a}{0}{0}, \; \matrixtt{a}{b}{a}{b},\; \mathrm{and} \; \matrixtt{a}{a}{b}{b}
\]

The physical meaning of a singular Jones matrix can be grasped by substituting these into Eq.~(\ref{eq:e-voltage}). The first two examples correspond to an antenna measuring zero voltage on one of the receptors (e.g. a broken wire). The latter two are examples of redundant measurements: both receptors will measure the same voltage, or linearly dependent voltages (consider, e.g., a flat aperture array, with a source in the plane of the dipoles). In all four cases there's irrecoverable loss of polarization information, so a polarization closure relation like Eq.~(\ref{eq:closure}) breaks down. (Note that the scalar analogue of this is simply a null scalar visibility, in which case scalar closures also break down.)

In the wide-field or all-sky case (Eq.~\ref{eq:me-allsky}), simple closures (whether matrix or scalar) no longer apply. However, the \emph{contribution} of each discrete point source to the overall visibility is still subject to a closure relationship. It is perhaps useful to formulate this in differential terms. Consider a brightness distribution $\coh{B}{}^{(0)}(\vec l)$, and let this correspond to a set of observed visibilities $\coh{V}{pq}^{(0)}$. Adding a point source of flux $\coh{B}{1}$ at position $\vec l_1$ gives us the brightness distribution:

\[
\coh{B}{}^{(1)}(\vec l) = \coh{B}{}^{(0)}(\vec l)+\delta(\vec l - \vec l_1)\coh{B}{1},
\]

where $\delta$ is the Kronecker delta-function, with corresponding observed visibilities $\coh{V}{pq}^{(1)}$. From the RIME (and Eq.~\ref{eq:me-allsky} in particular) it then necessarily follows that the \emph{differential} visibilities $\Delta\coh{V}{pq}=\coh{V}{pq}^{(1)} - \coh{V}{pq}^{(0)}$ will then satisfy the matrix closure relationship of Eq. (\ref{eq:closure}).

\section{Limitations of the RIME formalism\label{sec:rime-limitations}}

\subsection{\label{sec:noise}Noise}

The RIME as presented here and in the original papers is formulated for a noise-free measurement. In practice, each element of the $\coh{V}{pq}$ matrix (i.e. each complex visibility) is accompanied by uncorrelated Gaussian noise in the real and imaginary parts; a detailed treatment of this can be found in \citet[Sect.~6.2]{tms}. The noise level imposes a hard sensitivity limit on any given observation, which has a few implications relevant to our purposes:

\begin{itemize}
\item ``Reaching the noise'' has become the ``gold standard'' of calibration \citep[see Paper II,][]{RRIME2}. 
Many reductions are limited by calibration artifacts rather than the noise.
\item {\em Corrections} to the data (however one defines the term) can potentially distort the noise level across an observation in complicated ways, so due care must be taken.
\item Faint sources below the noise threshold can be effectively ignored.
\item Numerical approximations can be considered ``good enough'' once they get to within the noise (assuming no systematic errors), but see Paper III \citep[Sect.~2.6, Fig.~17]{RRIME3} for a big caveat to this.
\end{itemize}

The latter two considerations are what I refer to by ``sufficiently faint'' sources and ``sufficiently close'' approximations throughout this series of papers.

\subsection{\label{sec:smearing}Smearing and decoherence}

In Sect.~\ref{sec:RIME-emerges}, when going from Eq.~(\ref{eq:corr1}) to (\ref{eq:corr2}), we assumed that the Jones matrix $\jones{J}{p}$ is constant over the time/frequency bin of the correlator. That this is, strictly speaking, never actually the case can be seen from the definition of the $K$-Jones term in Eq.~(\ref{eq:K}). The vector $\vec u_p$ is defined in units of wavelength, making $K_p$ variable in frequency. The Earth's rotation causes $\vec u_p$ to rotate in our (fixed relative to the sky) coordinate frame, which also makes variable in time. To take this into account, the RIME (in any form) should be rewritten as an integration over a time/frequency interval. For example, the basic RIME of Eq.~(\ref{eq:me0}), when considering the integration bin $[t_0,t_1]\times[\nu_0,\nu_1]$, should be properly rewritten as:

\begin{eqnarray}
\langle \coh{V}{pq} \rangle & = & \frac{1}{\Delta t\Delta\nu}\int\limits^{t_1}_{t_0} \int\limits^{\nu_1}_{\nu_0} \coh{V}{pq}(t,\nu)\,d\nu\,dt \nonumber \\
\label{eq:me0:int}
& = & \frac{1}{\Delta t\Delta\nu}\int\limits^{t_1}_{t_0} \int\limits^{\nu_1}_{\nu_0} \jones{J}{p} (t,\nu) \coh{B}{}  \jonesT{J}{q}(t,\nu) \, d\nu\,dt,
\end{eqnarray}

which becomes Eq.~(\ref{eq:me0}) at the limit of $\Delta t,\Delta\nu \to 0$. Since $\jones{J}{}$ contains $K$, the complex phase of which is variable in frequency and time, the integration in Eq.~(\ref{eq:me0:int}) always results in a net loss of amplitude in the measured $\langle \coh{V}{pq} \rangle $. This mechanism is well-known in classical interferometry, and is commonly called {\em time/bandwidth decorrelation} or {\em smearing}. Note that a phase variation in any other Jones term in the signal chain will have a similar effect. The VLBI community knows of it in the guise of {\em decoherence} due to atmospheric phase variations; in RIME terms, atmospheric decoherence is just Eq.~(\ref{eq:me0:int}) applied to ionospheric $Z$-Jones or tropospheric $T$-Jones.\footnote{Small interferometers see very little atmospheric decoherence: if $Z_p\approx Z_q$ (as is the case for closely located stations), then $Z_p Z^\herm_q \approx 1$, so there is no net phase contribution to the integrand of Eq.~(\ref{eq:me0:int}).} I shall use the term {\em decoherence} for the general effect; and {\em smearing} for the specific case of decoherence caused by the $K$ term.

The mathematics of smearing are well-known for the scalar case, see e.g. \citet[Sect.~6.4]{tms} and \citet{Bridle:smearing}. Smearing increases with baseline length ($\vec u_{pq}$) and distance from phase center ($l,m$). Since the noise amplitude does {\em not} decrease, smearing results in a decrease of sensitivity. \citet{ME1} mention smearing in the context of the RIME. Since integration (and thus smearing) of a matrix equation is an element-by-element operation,  treatment of smearing within the RIME formalism is a trivial extension of the scalar equations.

For the general case of decoherence, a useful first-order approximation can be obtained by assuming that $\Delta t$ and $\Delta\nu$ are small enough that the amplitude of $\coh{V}{pq}$ remains constant, while the complex phase varies linearly. The relation

\[
\int\limits_{0}^{x_0}\mathrm{e}^{ix}dx = \mathrm{sinc}\frac{x_0}{2}e^{ix_0/2},
\]

which is well-known from the case of smearing with a square taper, then gives us an approximate equation for decoherence, in terms of the phase changes in time ($\Delta\vec\Psi$) and frequency ($\Delta\vec\Phi$):

\begin{eqnarray}\label{eq:smearing}
\langle \coh{V}{pq} \rangle & \simeq & \mathrm{sinc}\frac{\Delta\vec\Psi}{2}\,\mathrm{sinc}\frac{\Delta\vec\Phi}{2}\,\coh{V}{pq}(t_\mathrm{mid},\nu_\mathrm{mid}), \\
\nonumber && \mathrm{where} \; t_\mathrm{mid} = (t_0+t_1)/2, \nu_\mathrm{mid} = (\nu_0+\nu_1)/2, \\
\nonumber && \Delta\vec\Psi = \arg \coh{V}{pq}(t_1,\nu_\mathrm{mid}) - \arg \coh{V}{pq}(t_0,\nu_\mathrm{mid}), \\
\nonumber && \Delta\vec\Phi = \arg \coh{V}{pq}(t_\mathrm{mid},\nu_1) - \arg \coh{V}{pq}(t_\mathrm{mid},\nu_0) 
\end{eqnarray}

Equation (\ref{eq:smearing}) is straightforward to apply numerically, and is independent of the particular form of $\jones{J}{}$ responsible for the decoherence. However, the assumption of linearity in phase over the time/frequency bin can only hold for the visibility of a single source. In fact, it is easy to see that {\em any} approximation treating decoherence as an amplitude-only effect can, in principle, only apply on a source-by-source basis -- just consider the case of smearing, which varies significantly with distance from phase centre. In an equation like (\ref{eq:me-nps-ge}), the approximation can be applied to each term in the sum individually, or at least to as many of the brightest sources as is practical. This approach was used for the calibration described in Paper III \citep{RRIME3}.

\subsection{\label{sec:closure-errors}Interferometer-based errors}

The term {\em interferometer-based errors} refers to measurement errors that cannot be represented by per-antenna terms.
These are also called {\em closure errors}, since they violate the closure relationships of Sect.~\ref{sec:closures}. When formulating Eq.~(\ref{eq:me0}), we assumed that the visibility matrix $\coh{V}{pq}$ output by the correlator is a perfect measurement of correlations between antenna voltages. Closure errors represent additional baseline-based effects. Assuming these are linear, and following \citet{JEN:note185}, we could rewrite the full-sky RIME of Eq.~(\ref{eq:me-allsky-simple}) as: 

    \begin{equation}\label{eq:me:closure-errors}
    \coh{V}{pq} = \coh{M}{pq} \ast ( \jones{J}{p} \coh{X}{pq}  \jonesT{J}{q} ) + \coh{A}{pq},
    \end{equation}

where $\coh{M}{pq}$ is a $2\times2$ matrix of multiplicative interferometer errors, $\coh{A}{pq}$ is a $2\times2$ matrix of additive errors, and ``$\ast$'' represents element-by-element (rather than matrix) multiplication.

Given a model for $\coh{X}{pq}$, observed data $\coh{V}{pq}$, and self-calibrated per-antenna terms $\jones{J}{p}$, it is trivial to estimate $\coh{M}{}$ and $\coh{A}{}$ using Eq.~(\ref{eq:me:closure-errors}). It is also trivial to see that the equation is ill-conditioned: any model $\coh{X}{}$ can be made to fit the data by choosing suitable values for $\coh{M}{}$ and $\coh{A}{}$. We therefore need to assume some additional constraints, such as closure errors being fixed (or only slowly varying) in time and/or frequency. 

In practice, closure errors arise due to a combination of effects:

\begin{itemize}
\item The traditional ``purely instrumental'' cause is the use of analog components in the signal chain and parts of the correlator, which is typical of the previous generations of radio interferometers. New telescope designs tend to digitize the signal much closer to the receiver, and use all-digital correlators, presumably eliminating instrumental closure errors. 
\item Smearing and decoherence (Sect.~\ref{sec:smearing}) is a baseline-based effect, and will thus manifest itself as a closure errors, unless it is properly taken into account in the model for $\coh{X}{pq}$.
\item In general, any source structure or flux not represented by the model $\coh{X}{pq}$ will also show up as a closure error.
\end{itemize}

A solution for $\coh{M}{}$ and/or $\coh{A}{}$ will tend to subsume all these effects. This is dangerous, as it can actually attenuate sources in the final images, as illustrated in Paper III \citep[Sect.~1.5]{RRIME3}. One must thus be very conservative with closure error solutions, lest they become just another ``fudge factor'' in the equations.

\subsection{A three-dimensional RIME?\label{sec:3D-rime}}

Recent work by \citet{Carozzi:ME3D} highlights a limitation of the $2\times2$ Jones formalism. They point out that 
since we're measuring a 3D brightness distribution, the radiation from off-center sources is only approximately paraxial (equivalently, the EM waves are only approximately transverse). From this it follows that a 2D description of the EMF 
based on a rank-2  vector (the $\vec e$ used above) is insufficient, and a rank-3 formalism is proposed. 

The main implication of the Carozzi-Woan result for the $2\times2$ formalism is that the latter is still valid in general (at least for dual-receptor arrays), but the full-sky RIME of Eq.~(\ref{eq:me-allsky0}) must be augmented with an additional direction-dependent Jones term called the \emph{$xy$-projected transformation matrix}, designated as $\jones{T}{}^{(xy)}$ (see their Eq.~34), which corresponds to a projection of the 3D brightness distribution onto the plane of the receptors. If all the receptors of the array are plane-parallel (Carozzi \& Woan call this a {\em plane-polarized} interferometer), $\jones{T}{}^{(xy)}$ is a trivial DDE (in the sense of Eq.~\ref{eq:trivial-ddes}), manifesting itself as a polarization aberration that increases with $l,m$ (see their Fig.~2). For non-parallel receptors, $\jones{T}{}^{(xy)}$ should be a non-trivial DDE!

Classical dish arrays are plane-polarized by design, but deviate from this in practice due to pointing errors and other misalignments. The resulting effect is expected to be tiny given the typically narrow FoV of a dish, but it would be intriguing to see whether it can be detected in deliberately mispointed WSRT observations, given the extremely high dynamic range routinely achieved at the WSRT. On the other hand, an aperture array such as LOFAR should show a far more significant deviation from the plane-polarized case (due to the curvature of the Earth, as well as the all-sky FoV). With LOFAR's (as yet) relatively low dynamic range and extreme instrumental polarization, the effect may be challenging to detect at present. Further work on the subject is urgently required, given the polarization purity requirements of future telescopes (and in particular the SKA).

\section{\label{sec:formulations}Alternative formulations}

\subsection{Mueller vs. Jones formalism\label{sec:mueller}}

The original paper by \citet{ME1} formulated the RIME in terms of $4\times4$ {\em Mueller} matrices \citep{Muller}. This is mathematically fully equivalent to the $2\times2$ form introduced by \citet{ME4} in the fourth paper, and has since been adopted by many authors \citep{JEN:note185,tms,SB:imageplane,Rau:DDEs}. In my view, this is somewhat unfortunate, as the $2\times2$ formulation is both simpler and more elegant, and has far more intuitive appeal, especially for understanding calibration problems. For completeness, I will make an explicit link to the $4\times4$ form here.

Instead of taking the matrix product of two voltage vectors $\vec v_p$ and $\vec v_q$ and getting a $2\times2$ visibility matrix, as in Eq.~(\ref{eq:coherency}), we can take the {\em outer product} of the two to get the {\em visibility vector} $v_{pq}$:

\[
\vec v_{pq} = 2 \left< \vec v_p \otimes \vec v^\herm_q \right > = 2 \left ( 
\begin{array}{c}
    \langle v_{pa}v^*_{qa}\rangle \\ \langle v_{pa}v^*_{qb}\rangle \\
    \langle v_{pb}v^*_{qa}\rangle \\ \langle v_{pb}v^*_{qb}\rangle \\
\end{array} 
\right ) 
\]

Combining this with Eq.~(\ref{eq:e-voltage}), we get

\[
    \vec v_{pq} = 2 ( \jones{J}{p} \otimes \jonesT{J}{q} ) (\vec e \otimes \vec e^\herm )
 = ( \jones{J}{p} \otimes \jonesT{J}{q} ) 
\left ( \begin{array}{c}
I+Q \\ U+iV \\ U-iV \\ I-Q
\end{array} \right ), 
\]

which then gives us the $4\times4$ form of Eq.~(\ref{eq:me0}):

    \begin{equation}\label{eq:me:mueller}
    \vec v_{pq} = ( \jones{J}{p} \otimes \jonesT{J}{q} ) \jones{S}{} \jones{I}{} = {\cal J}_{pq} \jones{S}{} \jones{I}{}
    \end{equation}

Here, ${\cal J}_{pq}=\jones{J}{p} \otimes \jones{J}{q}$ is a $4\times4$ matrix describing the combined effect of the signal paths to antennas $p$ and $q$, $\jones{I}{}$ is a column vector of the Stokes parameters $(I,Q,U,V)$, and $\jones{S}{}$ is a conversion matrix that turns the Stokes vector into the brightness vector
\footnote{A Mueller matrix represents a linear operation on Stokes vectors, and so does not explicitly appear in these equations. For Eq.~(\ref{eq:me:mueller}), the equivalent Mueller matrix is  $\jonesinv{S}{}{\cal J}_{pq}\jones{S}{}$.}:

\[
\left ( \begin{array}{c}
I+Q \\ U+iV \\ U-iV \\ I-Q
\end{array} \right ) 
= \jones{S}{} 
\left ( \begin{array}{c}
I \\ Q \\ U \\ V
\end{array} \right ) 
\]

The equivalent of the ``onion'' form of Eq.~(\ref{eq:me0-onion}) is then:

    \begin{equation}\label{eq:me:mueller-onion}
    \vec v_{pq} = ( \jones{J}{pn} \otimes \jonesT{J}{qn} ) ... ( \jones{J}{p1} \otimes \jonesT{J}{q1} ) \jones{S}{} \jones{I}{}
= {\cal J}_{pqn} ...  {\cal J}_{pq1} \jones{S}{} \jones{I}{}
    \end{equation}

Likewise, the full-sky RIME of Eq.~(\ref{eq:me-allsky}) can be written in the $4\times4$ form as:

    \begin{equation}\label{eq:allsky:mueller}
\vec v_{pq} = {\cal G}_{pq} \iint\limits_{lm} {\cal E}_{pq}(l,m) \jones{S}{} \jones{I}{}(l,m) {\rm e}^{-2\pi i(u_{pq} l+v_{pq} m+w_{pq} (n-1))} \,dl\,dm 
    \end{equation}

This form of the RIME is particularly favoured when describing imaging problems \citep{SB:imageplane,Rau:DDEs}. It emphasizes that an interferometer performs a linear operation on the sky distribution $\jones{I}{}(l,m)$, via the linear operators ${\cal G}_{pq}$, ${\cal E}_{pq}(l,m)$, and the Fourier Transform $\cal F$, while eliding the internal structure of ${\cal G}$ and ${\cal E}$.

On the other hand, if we're interested in the underlying physics of signal propagation (as is often the case for calibration problems), then the $4\times4$ form of the RIME becomes extremely opaque. When considering any specific set of propagation effects (and its corresponding Jones chain), the outer product operation turns simple-looking $2\times2$ Jones matrices into an intractable sea of indices; see \citet[Eq. 4]{SB:imageplane} and \citet[Appendix A]{ME1} for typical examples. The $2\times2$ form provides a more transparent description of calibration problems, and for this reason is also far better suited to teaching the RIME. An excellent example of this transparency is given in Paper II \citep[Sect.~2.2.2]{RRIME2}, where I consider the effect of differential Faraday rotation.

There are also potential computational issues raised by the $4\times4$ formalism. A naive implementation of, e.g., Eq.~(\ref{eq:me:mueller-onion}) incurs a series of $4\times4$ matrix multiplications for each interferometer and time/frequency point. Multiplication of two $4\times4$ matrices costs 112 floating-point operations (flops), and the outer product operation another 16. Therefore, each pair of Jones terms in the chain incurs 128 flops. The same equation in $2\times2$ form invokes 12 floating-point operations (flops) per matrix multiplication, or 24 per each pair of Jones terms. This is roughly 5 times fewer than the $4\times4$ case. 

Often, the true computational bottleneck lies elsewhere, i.e. in solving (for calibration) or gridding (for imaging), in which case these considerations are irrelevant. However, when running massive simulations (that is, using the RIME to predict visibilities), my profiling of MeqTrees has often shown matrix multiplication to be the major consumer of CPU time. In this case, implementing calculations using the $2\times2$ form represents a significant optimization.

\subsection{Jones-specific formulations\label{sec:jones-specific}} 

Formulations of the RIME such as Eqs.~(\ref{eq:me-allsky}) or (\ref{eq:me-nps-ge}) are entirely general and non-specific, in the sense that they allow for any combination of propagation effects to be inserted in place of the $\jones{G}{}$ and $\jones{E}{}$ terms. A specific formulation may be obtained by inserting a particular sequence of Jones matrices. The first RIME paper \citep{ME1} already suggested a specific Jones chain. This was further elaborated on by \citet{JEN:note185}, and eventually implemented in AIPS++, which subsequently became CASA. The Jones chain used by current versions of CASA is described by \citet[Appendix E.1]{CASA:UserRef}:

\begin{equation}\label{eq:casa}
\jones{J}{p} = \jones{B}{p} \jones{G}{p} \jones{D}{p} \jones{E}{p} \jones{P}{p} \vec  T_p
\end{equation}

The Jones matrices given here correspond to particular effects in the signal chain, with specific parameterizations (e.g. $\jones{B}{p}$ is a frequency-variable bandpass, $\jones{G}{p}$ is time-variable receiver gain, etc.) Other authors \citep{Rau:DDEs} suggest variations on this theme. 

Such a ``Jones-specific'' approach has considerable merit, in that it shows how different real-life propagation effects fit together, and gives us {\em something} specific to be thought about and implemented in software. It does have a few pitfalls which should be pointed out.

The first pitfall of this approach is that it tends to place the trees firmly before the forest. A major virtue of the RIME is its elegance and simplicity, but this gets obscured as soon as elaborate chains of Jones matrices are written out.  I submit that the RIME's slow acceptance among astronomers at large is, in some part, due to the literature being full of equations similar to (\ref{eq:casa}). That they are just specific cases of what is at core a very simple and elegant equation is a point perhaps so obvious that some authors do not bother noting it, but it cannot be stressed enough!

The second pitfall is that an equation like (\ref{eq:casa}), when implemented in software, can be both too specific, and insufficiently flexible. (Note that the CASA implementation specifies both the time/frequency behaviour, and the form of the Jones terms, e.g. $\jones{G}{}$ is diagonal and variable in time, $\jones{B}{}$ is diagonal and variable in frequency,
$\jones{D}{}$ has a specific ``leakage'' form, etc.) For instance, the calibration described in Paper III \citep{RRIME3} cannot be done in CASA, despite using an ostensibly much simpler form of the RIME, because it includes a Jones term that was not anticipated in the CASA design. A second major virtue of the RIME is its ability to describe different propagation effects; this is immediately compromised if only a specific and limited set of these is chosen for implementation.

A final pitfall of the Jones-specific view is that it tends to stereotype approaches to calibration. Equation~(\ref{eq:casa}) is a huge improvement on the \emph{ad hoc} approaches of older software systems, but in the end it is just some model of an interferometer that happens to work well enough for ``classically-designed'' instruments such as the VLA and WSRT, in their most common regimes. It is {\bf not} universally true that polarization effects can be completely described by a direction-independent leakage matrix ($\jones{D}{p}$), or bandpass by $\jones{B}{p}$ -- it just happens to be a practical first-order model, which completely breaks down for a new instrument such as LOFAR, where e.g. ``leakage'' is strongly direction-dependent. In fact, even WSRT results can be improved by departing from this model, as Paper III \citep{RRIME3} will show. We must therefore take care that our thinking about calibration does not fall into a rut marked out by a specific series of Jones terms.

\subsection{\label{sec:circular}Circular vs. linear polarizations}

In Sect.~\ref{sec:derivation}, I mentioned that the RIME holds in any coordinate system. \citet{ME1} briefly 
discussed coordinate transforms in this context, but a few additional words on the subject are required.

Field vectors $\vec e$ and Jones matrices $\jones{J}{}$ may be represented [by a particular set of complex values] in any coordinate system, by picking a pair of complex basis vectors in the plane orthogonal to the direction of propagation. I have used an orthonormal $xy$ system until now. Another useful system is that of circular polarization coordinates $rl$, whose basis vectors (represented in the $xy$ system) are $\vec e_r=\frac{1}{\sqrt{2}}(1,-i)$ and $\vec e_l=\frac{1}{\sqrt{2}}(1,i)$. Any other pair of basis vectors may of course be used. In general, for any two coordinate systems S and T, there will be a corresponding $2\times2$ {\em conversion matrix} $\jones{T}{}$, such that $\vec e_\mathrm{T}=\jones{T}{} \vec e_\mathrm{S}$, where $\vec e_\mathrm{S}$ and $\vec e_\mathrm{T}$ represent the same vector in the S and T coordinate systems. Likewise, the representation of the linear operator $\jones{J}{}$ transforms as $\jones{J}{\mathrm{T}}=\jones{T}{} \jones{J}{\mathrm{S}} \jonesinv{T}{}$, while the brightness matrix $\coh{B}{}$ (or indeed any coherency matrix) transforms as $\coh{B}{\mathrm{T}}=\jones{T}{} \coh{B}{\mathrm{S}} \jonesT{T}{}.$

Of particular importance is the matrix for conversion from linear to circularly polarized coordinates. This matrix is commonly designated as $\jones{H}{}$ (being the mathematical equivalent of an electronic {\em hybrid} sometimes found in antenna receivers):

\[
\jones{H}{} = \frac{1}{\sqrt{2}} \matrixtt{1}{i}{1}{-i} \;\;\; \jonesinv{H}{} = \frac{1}{\sqrt{2}} \matrixtt{1}{1}{-i}{i}
\]

Consequently, the brightness matrix $\coh{B}{}$, when represented in circular polarization coordinates, has the following form (I'll use the indices ``$\odot$'' and ``$+$'' where necessary to disambiguate between circular and linear representations):

\[
\coh{B}{\odot} = \jones{H}{} \coh{B}{+} \jonesT{H}{} = \matrixtt{I+V}{Q+iU}{Q-iU}{I-V}
\]

While EMF vectors and Jones matrices may be represented using an arbitrary basis, the receptor voltages we actually measure are specific numbers. The voltage measurement process thus implies a {\em preferred} coordinate system, i.e. circular for circular receptors, and linear for linear receptors. 

It is of course possible to convert measured data into a different coordinate frame after the fact. It is also perfectly possible, and indeed may be desirable, to mix coordinate systems within the RIME, by inserting appropriate coordinate conversion matrices into the Jones chain. A commonly encountered assumption is that a ``VLA RIME'' must be written down in circular coordinates and a ``WSRT RIME'' in linear, but this is by no means a fundamental requirement! We're free to express part of the signal propagation chain in one coordinate frame, then insert conversion matrices at the appropriate place in the equation to switch to a different coordinate frame. In the onion form of the RIME (Eq.~\ref{eq:me0-onion}), this corresponds to a change of coordinate systems as we go from one layer of the onion to another. For example:

\[
\coh{V}{pq} = \jones{G}{p} \jones{H}{} \left ( \sum_{s} \jones{E}{sp} \coh{X}{spq} 
\jonesT{E}{sq} \right ) \jonesT{H}{} \jonesT{G}{q}
\] 

One reason to consider the use of mixed coordinate systems is the opportunity to optimize the representation of particular physical effects. As an example, a rotation in the $xy$ frame (e.g. ionospheric Faraday rotation, or parallactic angle) is represented by a diagonal matrix in the $rl$ frame. If the observed field has no intrinsic linear polarization, the $\coh{B}{\odot}$ matrix is also diagonal. If a part of the RIME is known to contain diagonal matrices only, their product can be evaluated with significant computational savings (compared to the full $2\times2$ matrix regime). On the other hand, if the instrument is using linear receptors, then receiver gains ($\jones{G}{}$) should be expressed in the linear frame, lest calibrating them become extremely awkward. We should therefore implement the RIME somewhat like the above equation, with the appropriate $\jones{H}{}$ matrices inserted as ``late'' in the chain as possible, so that only the minimum amount of computation is done for the full $2\times2$ case. This approach is not yet exploited by any existing software, but perhaps it should be. In particular, the MeqTrees system \citep{meqtrees} automatically optimizes internal calculations when only diagonal matrices are in play, and would provide a suitable vehicle for exploring this technique.

Note that the {\em configuration matrix} $\jones{C}{}$ proposed by \citet{ME1}, and further discussed by \citet{JEN:note185}, plays a similar role, in that it converts from ``antenna frame'' to ``voltage frame''. Here I simply suggest a generalization of this line of thinking. The RIME allows for an arbitrary mix of coordinate frames, as long as the appropriate conversion matrices are inserted in their rightful places.\footnote{Nor should we restrict our thinking to just the $xy$ and $rl$ frames. It could well be that the RIME of a future instrument will turn out to have a particularly elegant form in some other coordinate basis.}

\section{Errors and controversies\label{sec:controversies}}

For all its elegance, even the simplest version of the RIME (e.g. as formulated in Sect.~\ref{sec:RIME-emerges}) contains two points of confusion and controversy. The first has to do with the sign of the $iV$ term, and the second with the factors of 2 in the definition of $\coh{V}{pq}$ and $\coh{B}{}$.

\subsection{Sign of Stokes $V$}

The sign of Stokes $V$ has been a perennial source of confusion. The \citet{IAU74} definition specifies that $V$ is positive for right-hand circular polarization, but the literature is littered with papers adopting the opposite convention. Fortunately, major software packages such as AIPS and MIRIAD follow the IAU definition (though this 
has not always been the case for their early versions). As for the $iV$ term in the RIME, Papers I and II of the original series \citep{ME1,ME2} used the sign convention of Eq.~(\ref{eq:IQUV}). In Paper III of the series, \citet{ME3} then discussed the issue in detail, and showed that this convention is ``correct'' in the sense of following from the IAU definitions for Stokes $V$ and standard coordinate systems. However, in Paper IV, \citet{ME4} then used the opposite sign convention! In Paper V, \citet{ME5} noted the inconsistency, yet persisted in using the opposite convention. 

For this series, I adopt the correct sign convention of the original RIME Papers I through III, as per Eq.~(\ref{eq:IQUV}).

In practice, few radio astronomers concern themselves with circular polarisation, which is perhaps why the confusion has been allowed to fester. Unfortunately, this also means that in the rare cases when sign of $V$ is important, it
must be fastidiously checked each time!

\subsection{\label{sec:factor2}Factors of 2, or what is the unit response of an ideal interferometer?}

A far more insidious issue is the factor of $2$ in Eqs.~(\ref{eq:coherency}) and (\ref{eq:IQUV}). This has been the subject of a long-standing controversy both in the literature and in software. The definition of Stokes $I$ in terms of the complex amplitudes of the electric field is quite unambiguous \citep{tms,born-wolf}. In particular:

\[
I=\langle |e_x|^2\rangle  + \langle |e_y|^2\rangle, \;\;\;
Q=\langle |e_x|^2\rangle  - \langle |e_y|^2\rangle.
\]

This implies that a {\em unit} source of $I=1, Q=U=V=0$ corresponds to complex amplitudes of $\langle |e_x|^2\rangle =\langle |e_y|^2\rangle = 1/2$. What is less clear is how to relate this to the outputs of a correlator. That is, given an ideal interferometer and a unit source at the phase centre, what visibility matrix $\coh{V}{pq}$ should we expect to see? (In other words, what is the gain factor of an ideal interferometer?) This is something for which no unambiguous definition exists. Historically, two conventions have emerged:

\paragraph{Convention-$\scriptstyle 1/2$.} Unity correlations correspond to unity complex amplitudes, so a 1 Jy source produces correlations of 1/2 each: 

\[
\coh{V}{pq} = \matrixtt{\langle |e_x|^2\rangle }{0}{0}{\langle |e_y|^2\rangle } = \frac{1}{2}\matrixtt{1}{0}{0}{1}
\]

\paragraph{Convention-1.} Unity correlations correspond to unity Stokes $I$:

\[
\coh{V}{pq} = 2\matrixtt{\langle |e_x|^2\rangle }{0}{0}{\langle |e_x|^2\rangle } = \matrixtt{1}{0}{0}{1}
\]

Convention-$\scriptstyle 1/2$ is somewhat more pleasing to the purists, as it retains standard physical units for visibilities. This is the convention used throughout the RIME papers, beginning with \citet{ME1}, and also originally adopted in the MeqTrees system \citep{meqtrees}. However, Convention-1 is by far the more widespread, having been adopted by AIPS and other software systems, which has caused it to become entrenched in the minds of most radio astronomers.

The first edition of what is effectively the main reference work of radio interferometry, \citet*{tms1}, had a factor of 1/2 in the equations for interferometer response (Eq.~4.46), but omitted it in Table~4.47. (I conjecture that this table may in fact be the origin of Convention-1!) By the time of the second edition, Convention-1 was already widespread, and the authors responded by dropping the factor of 1/2 after Eq.~(4.29), noting that it was ``omitted and considered to be subsumed within the overall gain factor.'' \citep[see p. 102]{tms}. For better or for worse, this has irrevocably consecrated Convention-1 as the one to follow.

Ultimately, flux scales are tied to known calibrator sources, whose brightnesses are quite unambiguously defined in units of janskys. This means that in practice, the factor of 2 is indeed quietly subsumed into the gain calibration. Problems arise when data is moved between software packages that follow different conventions. For example, data calibrated with MeqTrees (formerly using  Convention-$\scriptstyle 1/2$) is kept in a Measurement Set (MS), yet the only tool available for making images from an MS is the AIPS++/CASA imager (Convention-1). This has often resulted in images with fluxes that were off by a factor of 2, so the MeqTrees project has recently switched to Convention-1.

In this paper, I have taken the difficult decision of breaking with the original formulations, and recasting the RIME using Convention-1. There remains the question of where to inject the requisite factor of 2. I have decided to do it ``on the inside'', by dropping the factor of 1/2 from the \citet{ME4} definition of the brightness matrix $\coh{B}{}$ (Eq.~\ref{eq:IQUV}). The alternative was to add a factor of 2 to the ``outside'' of the equation. The ``inside'' approach appears to have a number of practical advantages:

\begin{itemize}
\item $\coh{B}{}$ becomes unity for a unit (1 Jy unpolarized) source.
\item The coherency of a point source at the phase centre (Sect.~\ref{sec:coherency}) becomes equivalent to its brightness (and not one-half of its brightness).
\item In the ``onion'' form of the ME (Eq.~\ref{eq:me0-onion}), each successive layer of the onion corresponds to measurable visibilities, without needing to carry an explicit factor of 2 around.
\end{itemize}

\section{Conclusions}

Since its original formulation by \citet{ME1}, the Radio Interferometer Measurement Equation (RIME) has provided the mathematical underpinnings for novel calibration methods and algorithms. Besides its explanatory power, the RIME formalism can be wonderfully simple and intuitive; this fact has become somewhat obscured by the many different directions that it has been taken in. Several authors have developed approaches to the DDE problem based on the RIME, using different (but mathematically equivalent) versions of the formalism. This paper has attempted to reformulate these using one consistent $2\times2$ formalism, in preparation for follow-up papers (II and III) that will put it to work. Finally, a number of misunderstandings and controversies has inevitably accrued themselves to the RIME over the years. Some of these have been addressed here. It is hoped that this paper has gone some way to making the RIME simple again. 

\bibliographystyle{aa}

\bibliography{16082}

\end{document}